\documentclass[fleqn,useAMS,usenatbib]{mnras}
\usepackage{threeparttable}
\usepackage{amssymb}
\usepackage{bm}
\usepackage{graphicx}
\usepackage{CJKutf8}
\usepackage{url}
\usepackage{amsmath}

\title[Baryonic mass-size relation]{The scaling relation between baryonic mass and stellar disc size of morphologically late-type galaxies}

\author[P.-F. Wu]{Po-Feng~Wu \begin{CJK*}{UTF8}{bkai}(吳柏鋒)\end{CJK*}
	\\
	Max-Planck-Institut f\"{u}r Astronomie, K\"{o}nigstuhl 17, D-69117 Heidelberg, Germany }

\begin{document}

\maketitle

\begin{abstract}
Here I report the scaling relation between the baryonic masses and the scale lengths of stellar discs from $\sim 1000$ morphologically late-type galaxies. The baryonic mass-size relation is a single power-law $R_\ast \propto M_b^{0.38}$  across $\sim3$ orders of magnitude in baryonic mass. The scatter in size at fixed baryonic mass is nearly constant and there is essentially no outlier.
The baryonic mass-size relation provides a more fundamental description of the structure of the discs than the stellar mass-size relation. The slope and the scatter of the stellar mass-size relation can be understood in the context of the baryonic mass-size relation. 
For gas-rich galaxies, the stars is no longer a good tracer for the baryons. High baryonic mass, gas-rich galaxies appear to be much larger at fixed stellar mass because most of the baryonic content is gas. The stellar mass-size relation thus deviates from the power law baryonic relation and the scatter increases at the low stellar mass end. Those extremely gas-rich low-mass galaxies can be classified as Ultra Diffuse Galaxies based on the structure.

\end{abstract}	


\section{Introduction}

Despite of the complicate processes taking place during the formation, disc galaxies exhibit tight correlations among their luminosities, sizes, and rotation speed \citep{sal96,cou07}. In the standard hierarchical galaxy formation model with cold dark matter (CDM), galaxies form out of the dissipational collapsed gas within the cold dark matter halos. The structural and dynamical properties of galaxies reflect the properties of the parent dark halos \citep{mo98,vdb98,vdb00,nav00}. Empirical galaxy scaling relations thus provide stringent constraints on galaxy formation and evolution models.

One of the most extensively studied scaling relations is the correlation between the luminosities and the rotation velocities of disc galaxies, the Tully-Fisher relation \citep{tul77}. The relation was first identified using the B-band luminosity. Follow-up studies showed that the slope of the Tully-Fisher relation depends on the wavelength. Also, the scatter in luminosity at fixed velocity reduced at longer wavelengths \citep{aar80,bot87,ber94,tul00}. Furthermore, some low surface brightness, gas-rich galaxies appear to be extreme outliers; they rotate too fast for their luminosities or stellar masses \citep{per91,zwa95,mat98,mcg00}. These features can be understood in the context of an arguably more fundamental scaling relation between the baryonic mass and the velocity, the baryonic Tully-Fisher relation \citep{wal99,mcg00}. When the baryonic mass, the total mass from stars and gas, is used as the metric instead of the luminosity, the correlation holds as a power-law for at least $\sim 4$ order of magnitudes in mass. The Tully-Fisher relation holds because the luminosity works as a proxy for the mass, therefore, the slope and the scatter depend on the mass-to-light ratio. Gas-rich galaxies deviate from the Tully-Fisher relation because the stellar light and stellar mass are no longer good proxies for the total baryonic mass. 

The luminosities, or the stellar masses, of the stellar discs also tightly correlate with their sizes. The correlations are often referred as the mass-size or luminosity-size relation. On average, brighter or more massive discs are larger \citep{cho85,she03,fat10}. However, the luminosity-size relation or the stellar mass-size relation is not a single power-law. At the low luminosity or low mass regime, the slope of the relation becomes shallower and the scatter in the size at fixed luminosity of stellar mass increases \citep{she03}.
Recently, a large number of very low surface brightness galaxies, named Ultra Diffuse Galaxies (UDGs), are identified in nearby galaxy clusters \citep{vd15a,kod15,mun15,vdb16}. These galaxies are more than an order of magnitude larger than the average size of galaxies with similar luminosities. They are extreme outliers in the luminosity-size relation.  
Follow-up studies identified more UDGs outside clusters \citep{mer16,lei17} and their formation mechanisms are in hot debate.

The lessons learned from the Tully-Fisher relation urge us to examine the correlation between the baryonic mass and the sizes of galaxy discs. In this paper, I will demonstrate that when the baryonic mass is used as the metric, the relation between the mass and the size holds as a single power law across $\sim 3$ orders of magnitude in mass. The observed stellar mass and morphological dependence of the stellar mass-size relation and the formation of some UDGs can be understood under the context of the baryonic mass-size relation.
I describe the sample and the measurements in Section~2. The mass-size relation of late-type galaxies is presented in Section~3. Section~4 discusses the result and the implication. This paper adopts the Hubble constant $H_0 = 75$ km~s$^{-1}$Mpc$^{-1}$.

\section{Data}

\subsection{Galaxy sample}

The goal of this paper is to study the relation between the baryonic masses and the scale lengths of the stellar discs of late-type galaxies, therefore, galaxies in discussion need to have both surface photometry and H\,{\sc i} gas content available. I draw the galaxy samples from two sources. The first part is selected from the \textit{Cosmicflow-Spitzer} project \citep{sor12}. The original purpose of the \textit{Cosmicflow-Spitzer} is to derive distances of galaxies using the Tully-Fisher relation \citep{tul77} from the \textit{Spitzer} \citep{wer04} 3.6$\mu$m photometry and H\,{\sc i} linewidth. The second part is from \citet{wu15}, who measured the surface brightness profiles of $\sim500$ extragalactic H\,{\sc i} source from the Arecibo Legacy Fast ALFA \citep[ALFALFA;][]{hay11} survey in the Wide-field Infrared Explorer \citep[WISE;][]{wri10} 3.4$\mu$m and 4.5$\mu$m bands. The surface photometry and the H\,{\sc i} gas content are available in the Extragalactic Distance Database \footnote{\url{http://edd.ifa.hawaii.edu/}} \citep[EDD;][]{tul09}.

A few selection criteria are applied to select galaxies studied in this paper. First, for \textit{Spitzer-Cosmicflow} galaxies, only whose H\,{\sc i} fluxes are also available in the EDD are included in order to measure the total baryonic mass. Second, I only include galaxies with numerical morphological type $T \geq -2$ (S0 or later type) in the Hyperleda catalog \citep{mak14} to ensure that an exponential fit to the galaxy disc is reasonable. The last, galaxies with axis ratio $b/a < 0.35$ (inclination angle $i>73$) are excluded from analysis. These galaxies are highly inclined, therefore, their de-projected disc structure parameters are less certain. 

There are 501 and 629 galaxies fulfill the above criteria for the \textit{Spitzer-Cosmicflow} and the \citet{wu15} galaxies, respectively. I refer the two sets of galaxies as CF2 and Wu15 samples. For 44 galaxies presenting in both samples, I adopt the measurements from the Wu15 sample. In total, 1086 galaxies are analyzed in this study. 

\subsection{Galaxy masses and sizes}

I derived the stellar mass based on the \textit{Spitzer} 3.6$\mu$m and WISE 3.4$\mu$m fluxes. For the Wu15 sample, the ratio between the stellar mass and 3.4$\mu$m flux is determined by the calibration of \citet{clu14} using WISE [3.4]-[4.5] colour \citep[see details in][]{wu15}. For the CF2 sample, not all galaxies are observed in \textit{Spitzer} 4.5$\mu$m, therefore, I adopt a constant mass-to-light ratio $\Upsilon_{3.6\mu m}$. Several calibrations have shown that a constant $\Upsilon_{3.6\mu m}$ is a good approximation \citep{mei14,mcg14,que15}. In this paper, I determine the $\Upsilon_{3.6\mu m}$ from 44 galaxies presenting in both the CF2 and Wu15 samples. A $\Upsilon_{3.6\mu m} = 0.47 M_\odot/L_\odot$ yields statistically consistent stellar mass measurements between the two samples. 

\begin{figure*}
	\centering
	\includegraphics[width=0.9\textwidth]{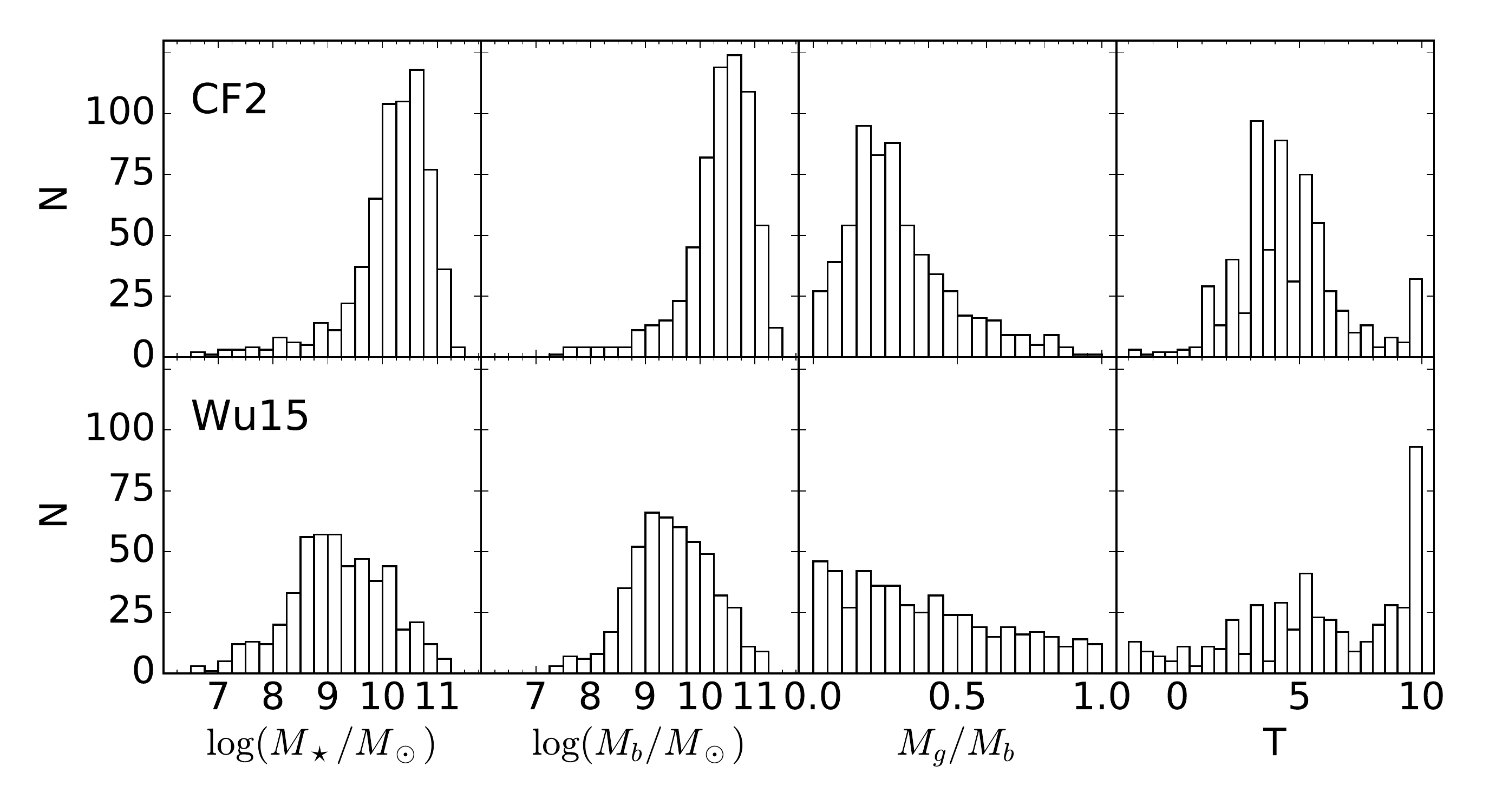}
	\caption{Distributions of galaxy properties. The upper and lower panels show the distribution of stellar masses, baryonic masses, gas fraction, and numerical morphological types (T) of the CF2 and the Wu15 sample, respectively. Galaxies in the Wu15 sample have on average lower masses and later morphological type. The Wu15 sample also contains more gas-rich galaxies.}
	\label{fig:dist}
\end{figure*}

The H\,{\sc i} mass is calculated as $M_{HI} = 2.356 \times 10^5 \times D^2 \times F$, where D is distance in Mpc and F is H\,{\sc i} flux in Jy~km~s$^{-1}$. I adopt the atomic gas mass $M_{atom} = 1.4 \times M_{HI}$ to include the contribution from helium and metals. There is no direct measurement for the molecular gas mass for most of galaxies. I estimate the molecular gas from the scaling relation formulate by \citet{wu15} basing on measurements of CO gas in $\sim 100$ galaxies \citep{bot14}: $\log(M_{H2}/M_{HI}) = 0.66 \times \log(M_\star/M_\odot) - 7.392$. 
For most galaxies, the molecular gas consists only a small fraction ($<10\%$) of the total baryonic content. Although the scatter in the CO gas mass among individual galaxies is not small \citep[$\sim0.4$~dex;][]{bot14}, the main conclusion of this paper is unlikely affected. I have also repeated the analysis without the molecular gas, i.e., only considering the stellar and H\,{\sc i} gas masses. The result remains the same. The baryonic mass in this paper refers to the sum of stellar mass, atomic gas mass, and molecular gas mass. The distributions of galaxy properties are shown in Fig.~\ref{fig:dist}.

In this paper, I use the disc scale length $R_d$ for the size. I refer readers \citet{sor12} and \citet{wu15} for detailed processes of the measurements. The photometry and structural parameters are available on EDD. Here I explain the salient points of the derivation.

The 1-D surface brightness profile is fit by an exponential form within a certain radius range.  The fiducial inner fitting range is set as the effective radius, the radius enclosing $50\%$ of stellar light. The outer fitting range is set as the isophot of $\mu_{3.6\mu m} =25.5$~mag~arcsec$^{-2}$ and $\mu_{3.4\mu m} = 25.0$~mag~arcsec$^{-2}$ for the CF2 and the Wu15 samples, respectively. This procedure mitigates the effect of the bar and the bulge, which could bias the measurement of scale length. I also measure the scale length with different inner fitting ranges, from the radius enclosing 25\% to 75\% of stellar light, to test the robustness of the measurement. I find that the derived scale lengths change by $\sim10\%$, where a smaller inner fitting range leads to on average shorter scale lengths. This systematics is similar to previous studies using SDSS optical bands \citep{fat10}. The main conclusion of this paper is not affected by the fitting range. Among 44 galaxies presenting in both samples, I find no systematic difference and a scatter of $\sim0.15$~dex between scale lengths measured from \textit{Spitzer} 3.6$\mu$m and WISE 3.4$\mu$m.

Another commonly used definition of the size of galaxies is the effective radius ($R_e$). For a pure exponential disc, $R_e = 1.68 R_d$. In general, the relation between $R_d$ and $R_e$ depends on the structure of the galaxy. I take $R_d$ as the definition of the size of the stellar disc and note it as $R_\ast$ hereafter. The main conclusion of this paper does not change if I use $R_e$ as the definition. Relevant figures are presented in the Appendix.

\section{Baryonic Mass--Size Relations}

\begin{figure*}
	\centering
	\includegraphics[width=0.9\textwidth]{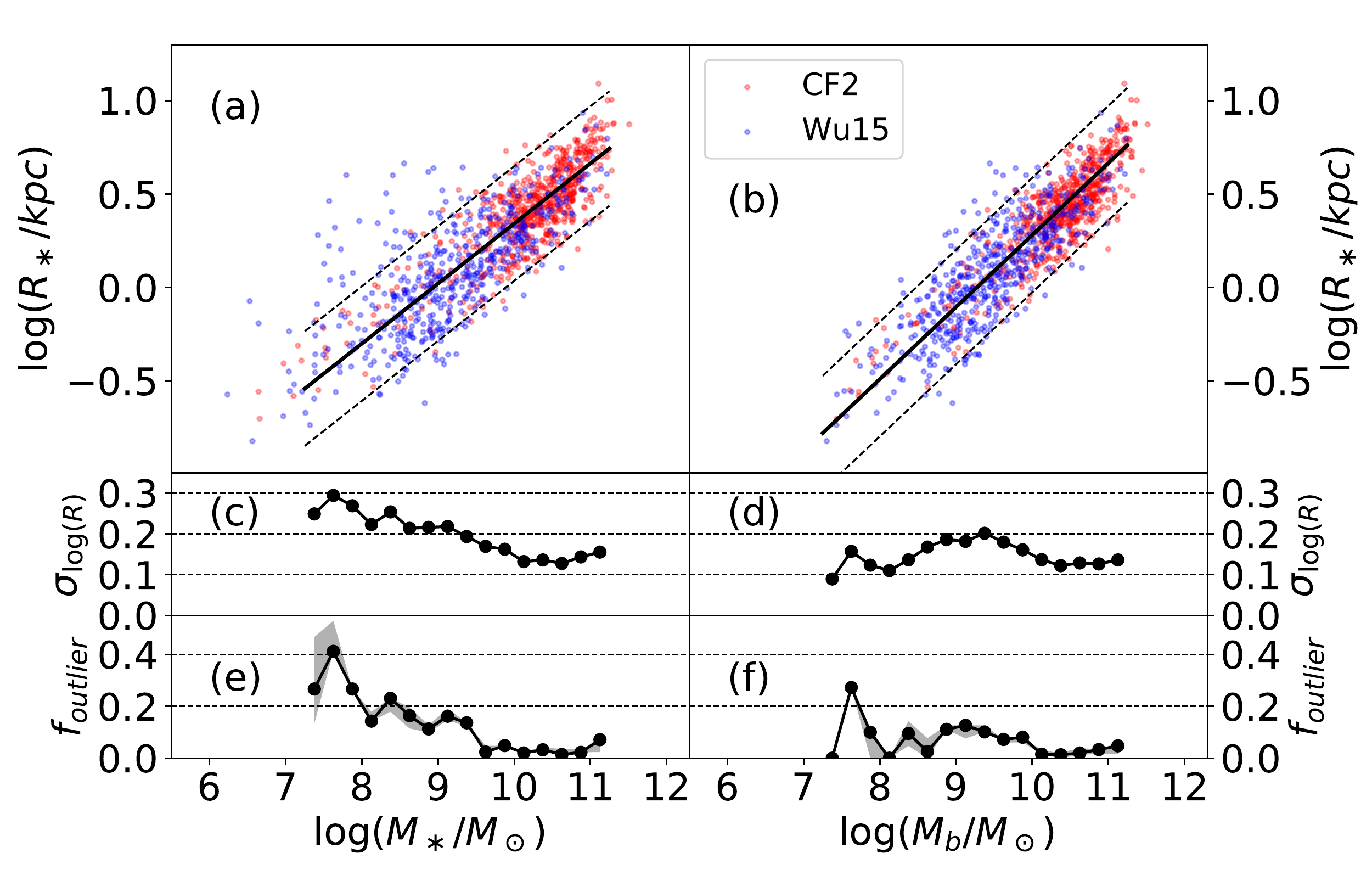}
	\caption{\textit{(a,b)} The sizes as a function of the stellar masses and the baryonic masses. The best-fit relations are plotted as the solid black lines. The dashed lines indicate 0.36~dex shifts in $R_\ast$. Galaxies from the CF2 and the Wu15 are labeled as red and blue dots, respectively. At lower stellar masses, several galaxies from the Wu15 sample appear to be much larger than the average at fixed stellar mass, while almost all galaxies follow the same scaling relation between the size and the baryonic mass. \textit{(c,d)} The scatter in $\log(R_\ast)$ at fixed masses. The scatter increases towards lower stellar masses. On the contrary, it is nearly constant at all baryonic masses. \textit{(e,f)} The fraction of outliers. The outliers is defined as galaxies which lie $>0.36$~dex (2$\sigma$ of the $R_\ast$--$M_b$ relation) away from the best-fit relation at fixed masses. The gray area shows the ranges of the outlier fraction if the best-fit relations are altered by 2$\sigma$.
	The fraction of outliers increases at $\log(M_\ast/M_\odot)\lesssim9$ but is generally $<0.1$ at all baryonic masses. }
	\label{fig:RM}
\end{figure*}

Fig.~\ref{fig:RM}a plots the size of the stellar disc $R_\ast$ as a function of stellar mass $M_\ast$. The CF2 and Wu15 samples are in red and blue, respectively. The tight correlation between $R_\ast$ and $M_\ast$ has been observed in previous studies \citep{cou07,fat10}. However, at the low $M_\ast$ regime, the number of galaxies deviating from the mean relation increases. Many galaxies with $M_\ast \simeq 10^8 M_\odot$ have stellar discs as large as discs in galaxies of $M_\ast \simeq 10^{10} M_\odot$. Fig.~\ref{fig:RM}c shows the scatter in sizes at fixed $M_\ast$. The scatter is calculated as the standard deviation of $\log(R_\ast)$ relative to the median in each mass bin. The scatter in sizes increases at low $M_\ast$, as seen by previous studies \citep{she03}. 
Fig.~\ref{fig:RM}a also shows that the outliers are all from the Wu15 sample, which is selected based on their H\,{\sc i} flux. Some galaxies in the CF2 sample also have stellar masses as low as $M_\ast \simeq 10^7 M_\odot$, but none of them deviates significantly from the mean $R_\ast$--$M_\ast$ relation. 

Fig.~\ref{fig:RM}b plots $R_\ast$ as a function of the baryonic mass ($M_b$). A tighter power-law relation holds between $R_\ast$ and $M_b$ for the entire mass range, with a scatter $\sigma_{\log(R)}\simeq0.18$~dex (Fig.~\ref{fig:RM}d). 
In addition, there is essentially no outliers in the $R_\ast$--$M_b$ relation. By using the baryons instead of stars as the metric for masses of galaxies, the relation between the size and mass of disc galaxies can be described as a single power-law down to at least $M_b \simeq 10^8 M_\odot$. 

I fit $\log(R_\ast)$ and $\log(M_b)$ to a linear relation using orthogonal distance regression. To estimate the uncertainty of the fit, I repeat the fit for 1000 times, for each time the mass and the size of each galaxy are disturbed by a small amount drawn from a Gaussian distribution, whose standard deviation is the uncertainty of the measurement.
The 50th percentile and the 16th and the 84th percentile from the 1000 trials are taken as the best-fit parameters and the uncertainties, respectively. I assign a conservative 0.35~dex uncertainty to the stellar mass according to the calibration of \citet{clu14}, but note that a much smaller uncertainty of 0.1~dex is also advocated by \citet{mei14}. The uncertainty in the molecular gas mass is set to be 0.4~dex \citep{bot14}. The uncertainty on the atomic gas mass comes from the uncertainty of the H\,{\sc i} 21~cm flux \citep{tul09,hay11}. For $R_\ast$, a 0.15~dex uncertainty is assigned to each galaxy (see Section~2). 
An extra uncertainty of the distance of 20\% \citep{tul13} is also included in the error budget. The best-fit relation for Fig~\ref{fig:RM}b is: 

\begin{equation}
\label{eq:mbrd}
\log(R_\ast/kpc) = (0.385^{+0.008}_{-0.013}) \times [\log(M_b/M_\odot)-10] + (0.281^{+0.010}_{-0.009})
\end{equation}
The slope and the scatter are in broad agreement with those of the stellar mass-size relation at high mass found in previous studies \citep[$\sim 0.3$--$0.4$, ][]{she03,cou07,fer13,mos13}. The best-fit $R_\ast$--$M_\ast$ relation has a shallower slope of 0.321$^{+0.011}_{-0.011}$ and an intercept of 0.343$^{+0.009}_{-0.009}$. 

Fig.~\ref{fig:RM}e and Fig.~\ref{fig:RM}f show the fraction of outliers at fixed $M_\ast$ and $M_b$, respectively. An outlier is defined as a galaxy whose scale length is larger or smaller than the average at fixed mass by 0.36~dex (2$\sigma$ of the $R_\ast$--$M_b$ relation). At fixed $M_\ast$, the fraction of outliers increases towards the low mass end. The outlier fraction is $\sim10\%$ at $\log(M_\ast/M_\odot)\simeq9$ and reaches $\sim30\%$ at $\log(M_\ast/M_\odot)\simeq8$. On the contrary, the $2\sigma$ outlier fraction at fixed $M_b$ is generally below 10\% at all masses, close to the 5\% expected from a normal distribution.

\begin{figure*}
	\centering
	\includegraphics[width=0.9\textwidth]{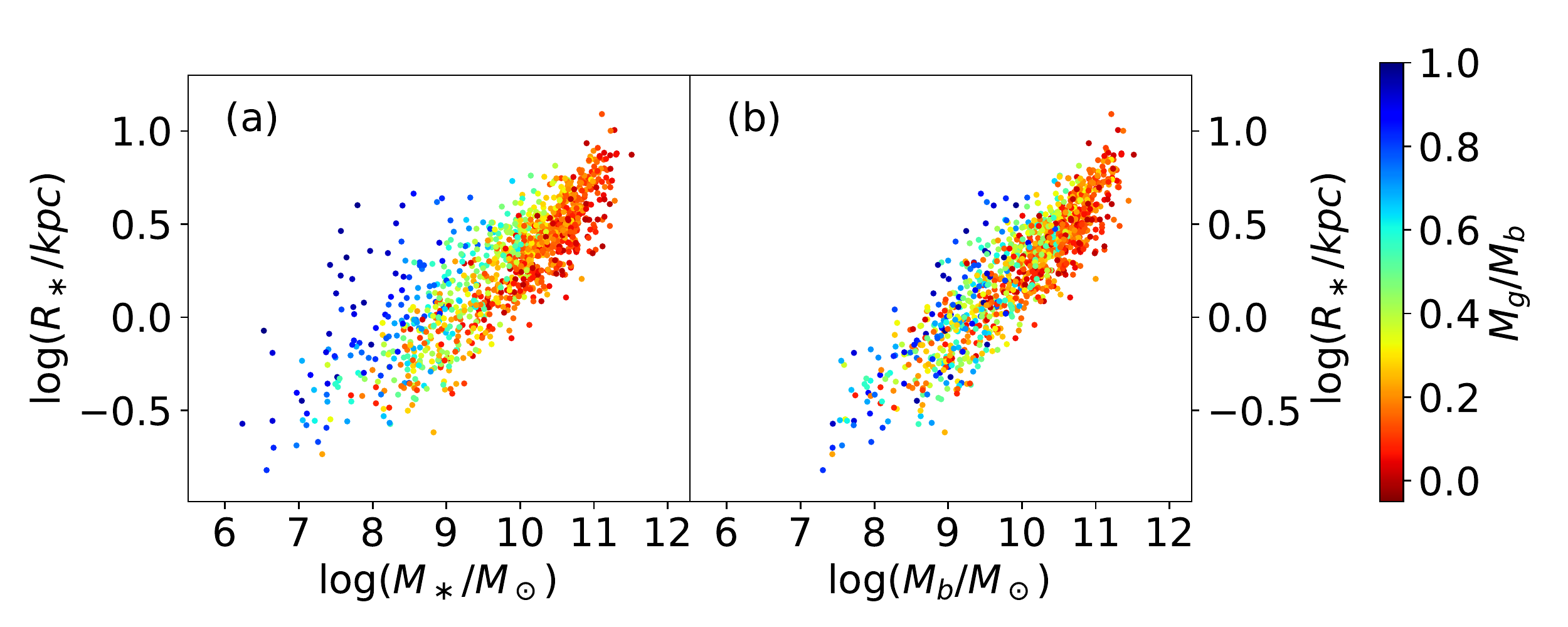}
	\caption{The relation between sizes and masses of galaxies, colour coded by the gas fraction. Similar to Fig.~\ref{fig:RM}, this figure shows the relation between the scale length and the stellar mass (left panel) and baryonic mass (right panel). The colour represents for the gas fraction, $M_g/M_b$, where gas-rich galaxies are in blue and gas-poor galaxies are in red. At fixed stellar mass, outliers from the mean relation are extremely gas-rich, with $M_g/M_b \sim 1$. The outliers in the stellar mass-size relation follow the average baryonic mass-size relation. Their sizes are comparable to other galaxies at fixed baryonic masses.}
	\label{fig:RMfg}
\end{figure*}

Fig.~\ref{fig:RMfg} shows again the $R_\ast$ as functions of $M_\ast$ and $M_b$, but the galaxies are colour-coded by their gas fraction ($M_g/M_b$). Fig.~\ref{fig:RMfg} shows that the outliers in the $R_\ast$--$M_\ast$ relation are galaxies with high gas fractions ($M_g/M_b \simeq 1$). While they follow the average $R_\ast$--$M_b$ relation, they appear to be much larger than galaxies of similar $M_\ast$ because most of their baryons are in the form of gas.

\begin{figure*}
	\centering
	\includegraphics[width=0.9\textwidth]{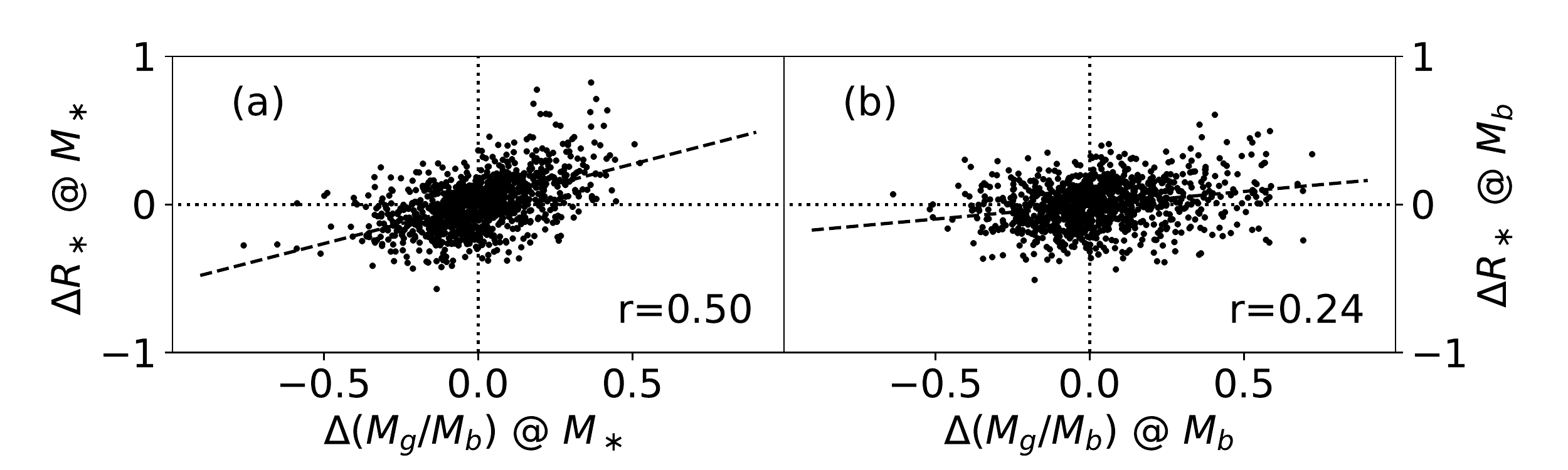}
	\caption{The correlation between the gas fraction and the sizes of galaxies. The $\Delta (M_g/M_b)$ and $\Delta R_\ast$ are the relative gas fraction and size of each galaxy comparing to galaxies of similar masses. Figure (a) and (b) plots the  $\Delta (M_g/M_b)$ and the $\Delta R_\ast$ calculated at fixed stellar masses and baryonic masses, respectively. The dashed lines are the best linear fits. The Spearman correlation coefficient is shown at the bottom-right of each panel. At fixed stellar mass, the size has an intermediate dependence on the gas fraction, where gas-rich galaxies are on average larger. The correlation is weaker at fixed baryonic mass. }
	\label{fig:dfgdr}
\end{figure*}

The correlation between the gas fraction and the galaxy size is further illustrated in Fig.~\ref{fig:dfgdr}. For each galaxy, I calculate the median size and the median gas fraction of galaxies of similar masses ($\Delta M < 0.1$~dex). The relative size ($\Delta R_\ast$) and gas fraction ($\Delta M_g/M_b$) of the galaxy are the difference between the value of the galaxy and the median. Fig.~\ref{fig:dfgdr} plots  $\Delta R_\ast$ as a function of $\Delta M_g/M_b$, calculated at fixed $M_\ast$ and $M_b$, respectively. On average, gas-rich galaxies are larger at fixed $M_\ast$. The Spearman correlation coefficient suggests a moderate correlation between the size and the gas fraction of galaxies at fixed $M_\ast$. On the other hand, the size depends only weakly on the gas fraction at fix $M_b$.

\section{Discussion}

\subsection{The scatter and the slope of the $R_\ast$-$M_\ast$ relation}

The $R_\ast$-$M_b$ relation is a single power law across $\sim3$ orders of magnitude in $M_b$. In this study, there is essentially no outliers and the scatter in $R_\ast$ does not change with $M_b$. 
On the contrary, the slope and the scatter of the $R_\ast$-$M_\ast$ relation depend on $M_\ast$ \citep{she03,dut12}.

This mass dependence of the $R_\ast$-$M_\ast$ relation can be understood in the context of the $R_\ast$-$M_b$ relation. Fig.~\ref{fig:RMfg} has shown that in general, the gas fraction, $M_g/M_b$, increases progressively toward the low-mass end. This result is consistent with H\,{\sc i} surveys on nearby galaxies \citep{mcg97,hua12,bot14,pee14}. At the high-mass end, the stellar mass is the dominant component, therefore, the slope and the scatter of the $R_\ast$-$M_\ast$ relation is similar to the $M_\ast$-$M_b$ relation \citep{she03,cou07,fer13,mos13}. 
However, at the low-mass end, the stellar mass is no long a good proxy for the total baryons. At fixed $M_\ast$, the scatter in size is a combination of the scatter in the $R_\ast$-$M_b$ relation and the scatter in the gas fraction. 
As a result, the scatter increases towards low $M_\ast$ end due to a larger variation in the gas masses and the baryonic masses of galaxies. The slope flattens at low $M_\ast$ because there is increasing number of gas-rich, high $M_b$ galaxies at given $M_\ast$, making the average size larger. 

Beyond the local Universe, \citet{vdw14} measured the stellar mass-size relation of star-forming galaxies up to $z = 3$ and found shallow slopes of $\sim0.22$ at all redshifts. At higher redshifts, star-forming galaxies are expected to have more gas than galaxies in the local Universe and may consist of a non-negligible fraction of the total baryonic budget \citep{dav11,fu12,tac13}. 
Therefore, the slope of $R_\ast$-$M_\ast$ relations may not reflect the slope of the $R_\ast$-$M_b$ relation even for massive star-forming galaxies. Future instruments like the Square Kilometer Array (SKA) will be needed for studying the evolution of the $R_\ast$-$M_b$ relation.

\subsection{Implication on Ultra Diffuse Galaxies}

The Ultra Diffuse Galaxies (UDGs) is a new category of galaxies which have low stellar masses ($M_\ast \lesssim 10^8 M_\odot$) but are 100 to 1000 times larger than galaxies of similar stellar masses. A large population of UDGs are first found in the Coma cluster \citep{vd15a,kod15}, with red optical colour and absent of star-formation activities \citep{vd15a,vd15b}. Later searches identified UDGs with similar structural properties but with a wide range of environments and optical colours \citep{mun15,mer16,yag16}. 

In Fig.~\ref{fig:RMfg}a, those extreme outliers would be classified as UDGs based on their structures. However, they follow the baryonic mass-size relation as other galaxies. There are another $\sim$100 similar galaxies also identified in the ALFALFA survey. These gas-rich UDGs are star-forming, with blue optical colours, and with high spin parameters \citep{lei17}. The high angular momenta prevent gas from collapsing and forming stars quickly, therefore, these galaxies remain gas-rich and appear to be outliers in the stellar mass-size relation \citep{amo16}. 

However, this formation mechanism may not necessarily apply to all UDGs. Those UDGs with red optical colours may be absent of star-forming activities, thus, likely gas-poor. In this case, the gas-poor UDGs will still be outliers in the $R_\ast$--$M_b$ plane (Fig~\ref{fig:RMfg}b) and appear to have less baryonic mass compared to galaxies of similar sizes. It is interesting that redder UDGs are on average located in denser regions \citep{vd15a,lei17}. The correlation between the environments and the colours may suggest a different route to form UDGs in clusters. The cluster UDGs may be galaxies which formed the first generation of stars then lost their gas through cluster-related mechanisms \citep{vd15a}. In this scenario, we may expect more UDGs in clusters than in less dense environments, which is consistent with recent observations \citep{vdb17}. Obtaining the gas mass of cluster UDGs and locating them on the $R_\ast$--$M_b$ plane will directly test the hypothesis.

\subsection{The scatter of the $R_\ast$-$M_b$ relation}
\label{sec:scatter}

At fixed mass, the size of discs is largely determined by its angular momentum content. It is usually assumed that the specific angular momentum of the disc correlates tightly with it of the parent halo, usually represented by the dimensionless spin parameter $\lambda$ \citep{pee69,bul01}, therefore, the scatter in the size-mass relation is determined by the intrinsic distribution of $\lambda$. Theoretical studies generally found that the distribution of $\lambda$ can be approximated by a log-normal function and the dispersion is nearly independent of the halo mass \citep{bet07,mac07,mac08}. Most studies found a dispersion $\sigma(\ln \lambda) \simeq 0.5$, which is $\sim 0.2$~dex. In the $R_\ast$-$M_b$ relation, the scatter in $R_\ast$ at fixed $M_b$ is $\sim0.18$~dex. Taking into account the $\sim0.15$~dex uncertainty in size measurement, the intrinsic scatter is $\sim 0.1$~dex, only half of the dispersion of $\lambda$ predicted for halos. The scatter in the size of late-type galaxies thus cannot be simply attributed to the spin of the haloes. Previous studies had also pointed out the smaller scatter in the $R_\ast$-$M_\ast$ relation and have not yet reached a consensus on its origin \citep{dut07,gne07,sai11,des15}.

The assumption of the strong correlation between the angular momentum of baryons and haloes is challenged by numerical simulations.
Several works have shown that mergers and feedback in star-formation processes can significantly alter the angular momentum of gas and haloes, resulting distinct distributions of spin parameter and misaligned angular momentum vectors between the gas and haloes \citep{che03,sha05,hah10,tek15,zju17}. Furthermore, the sizes are measured from the stellar component, which may not be representative for the baryonic disc. It is well known that the gas and the stars in disc galaxies do not have the same structure, where the extent of the gas is a few times larger than that of the stars \citep{oli91,mar95,vdb01,ton06,dut11,big12,kra13,obr14,but17}. For gas-rich galaxies, the sizes of stellar discs can be very different from the baryonic discs. It is intriguing that the scatter of the $R_\ast$-$M_b$ relation does not appear to be mass-dependent even for the mass range that the gas is on average the dominant component ($M_b \lesssim 10^9 M_\odot$). Moreover, a subset of the sample are extremely gas-rich, whose stellar components consist only $\sim10\%$ of the baryons but they still follow the same $R_\ast$-$M_b$ relation. It appears that detailed baryonic physics and star-formation processes need to be taken into account to explain the small scatter of the $R_\ast$-$M_b$ relation.

\subsection{The slope of the $R_\ast$-$M_b$ relation}

The size of the stellar disc and the baryonic mass scale as a single power law, $R_\ast \propto M_b^{0.385}$. For a virialized halo, the size and the mass are also expected to scale as a power law but with a different slope, $R_h \propto M_h^{1/3}$. The two scaling relations provide an opportunity to link the sizes of stellar discs with the sizes of dark halos. 

In order to do so, the first step is finding the mapping between the halo mass $M_h$ and the baryonic mass $M_b$. 
Studies using different methods have shown that the baryon-to-halo mass ratio ($M_b/M_h$) depends on halo mass, where for galaxy-size haloes, less massive halos retain smaller fractions of baryons \citep{dut12,pap12,san16}. Both \citet{dut12} and \citet{san16} provide the $M_b$ and $M_h$ of their model galaxies and their $M_b-M_h$ relations are in good agreement. For the mass range discussed in this paper, I approximate the $M_b-M_h$ relation based on the model galaxies in \citet{san16}:
\begin{equation}
\label{eq:mbmh}
\log(M_b/M_\odot) = 1.50 [\log(M_h/M_\odot)-11.50] + 10.00.
\end{equation}

For the relation between the size and the mass ofvirialized dark matter haloes:
\begin{align}
\label{eq:rhmh}
\begin{split}
R_h(kpc) & = \left(\frac{M_h G}{100H_0^2}\right)^{1/3} \\
         & = 134 \times \left(\frac{M_h}{10^{11.5}M_\odot}\right)^{1/3}.
\end{split}
\end{align}
Here the size and mass are defined as $R_{200}$ and $M_{200}$, the radius within which the mean mass density is 200 times of the critical density, and the mass within $R_{200}$, respectively.

From the observed $R_\ast-M_b$ relation, with the assumption of virialized haloes and the adopted $M_b-M_h$ relation (Equation~\ref{eq:mbrd},\ref{eq:mbmh}, and \ref{eq:rhmh}), the relation between the sizes of stellar discs and the haloes is:
\begin{equation}
\label{eq:rsrh}
R_\ast/R_h = 0.014 \left(\frac{M_h}{10^{11.5}M_\odot}\right)^{0.24}.
\end{equation}

The mass-dependence in Equation~\ref{eq:rsrh} is in apparent conflict with the result based on abundance matching. \citet{kra13} established the connection between haloes in simulation and observed galaxies by matching their cumulative halo and stellar mass functions. Based on this abundance matching ansatz, \citet{kra13} found that $R_\ast/R_h$ is a constant across 8 orders of magnitude in stellar mass.

However, Fig.~\ref{fig:RMfg}a and Fig.~\ref{fig:dfgdr}a have shown that at fixed $M_\ast$, larger galaxies contain on average more gas, thus, more massive in terms of baryons. One will expect that at fixed $M_\ast$, larger galaxies reside on average in more massive haloes. This positive correlation between halo masses and galaxy sizes at fixed stellar mass has been shown in both observations using halo masses from weak lensing and cosmological hydrodynamical simulations \citep{cha17}. 
As a result, the abundance matching based on the star-to-halo mass ratio is subject to the systematic dependence on galaxy sizes and the inferred disc-to-halo size ratio need to be examined with care \citep{kra13,hua17,som17}.

\section{Conclusion}
From $\sim1000$ local morphologically late-type galaxies, I demonstrate that the scaling relation between the scale lengths of the stellar discs and the baryonic masses is a single power law scaling as $R_\ast \propto M_b^{0.385}$ across $\sim3$ orders of magnitude in baryonic mass. The scatter in the scale lengths at fixed baryonic mass is independent of mass and there is essentially no outlier. The intrinsic scatter of the $R_\ast$-$M_b$ relation at fixed $M_b$ is $\sim0.1$~dex. This tight size distribution cannot be directly attributed to the halo spin, which would predict a wider size distribution.

The slope and the scatter of the $R_\ast$-$M_\ast$ relation can be understood in the context of the $R_\ast$-$M_b$ relation. At the low $M_\ast$ end, the star is no longer a good tracer for baryons. The scatter increases and the slope deviates from the $R_\ast$-$M_b$ relation because of the large variation of gas masses and baryonic masses in galaxies of similar stellar masses. 
Also, the disc sizes at fixed $M_\ast$ has a positive correlation with the gas fraction that larger galaxies contain on average more gas and baryons. This correlation suggests that abundance matching based on the star-to-halo mass relation is subject to the systematic dependence on galaxy size. 
	
The formation of gas-rich UDGs can be naturally explained by the $R_\ast$-$M_b$ relation. These gas-rich UDGs are likely galaxies with high angular momentum, which prevents gas from forming stars quickly. As a result, they appear to be outliers in the $R_\ast$-$M_\ast$ relation. However, some UDGs residing in galaxy clusters may form through losing gas due to cluster-related mechanisms. Measuring their gas mass and comparing them with the average $R_\ast$-$M_b$ relation will test this hypothesis.

\appendix
\section{The scaling relations using the effective radius}

\begin{figure*}
	\centering
	\includegraphics[width=0.9\textwidth]{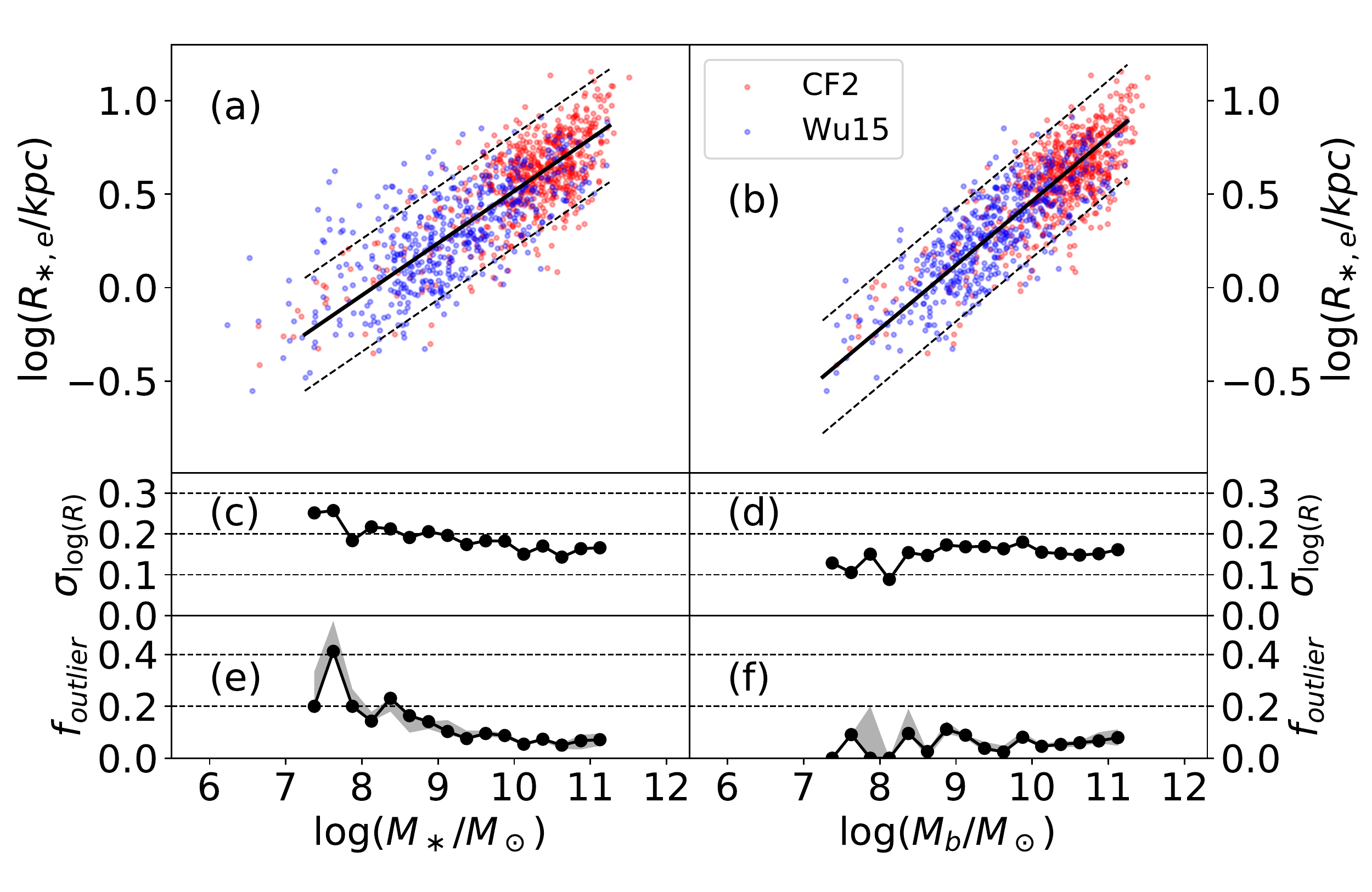}
	\caption{The effective radius, scatter, and outlier fraction as a function of the stellar mass and the baryonic mass. Same as Fig.~\ref{fig:RM}}
	\label{fig:ReM}
\end{figure*}

\begin{figure*}
	\centering
	\includegraphics[width=0.9\textwidth]{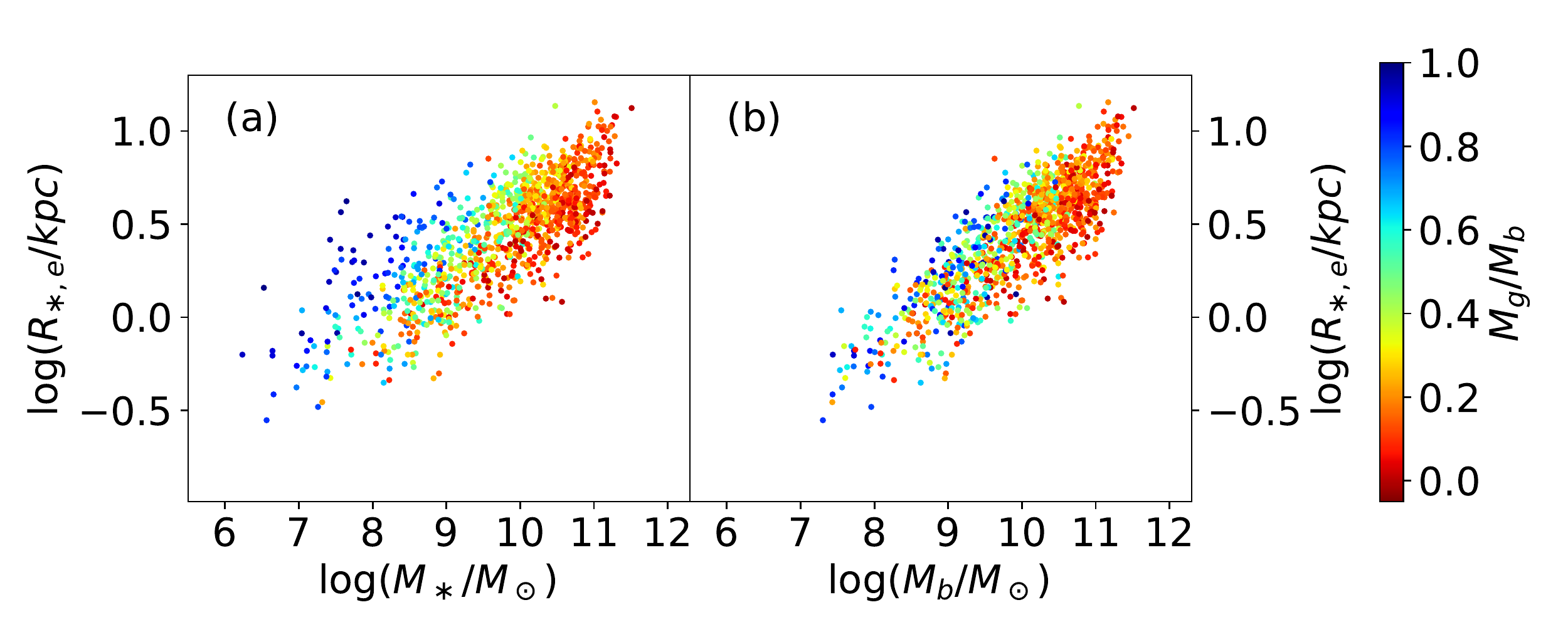}
	\caption{The relation between the effective radius and the stellar masses and the baryonic masses of galaxies, colour-coded by the gas fraction. Same as Fig.~\ref{fig:RMfg} }
	\label{fig:ReMfg}
\end{figure*}

\begin{figure*}
	\centering
	\includegraphics[width=0.9\textwidth]{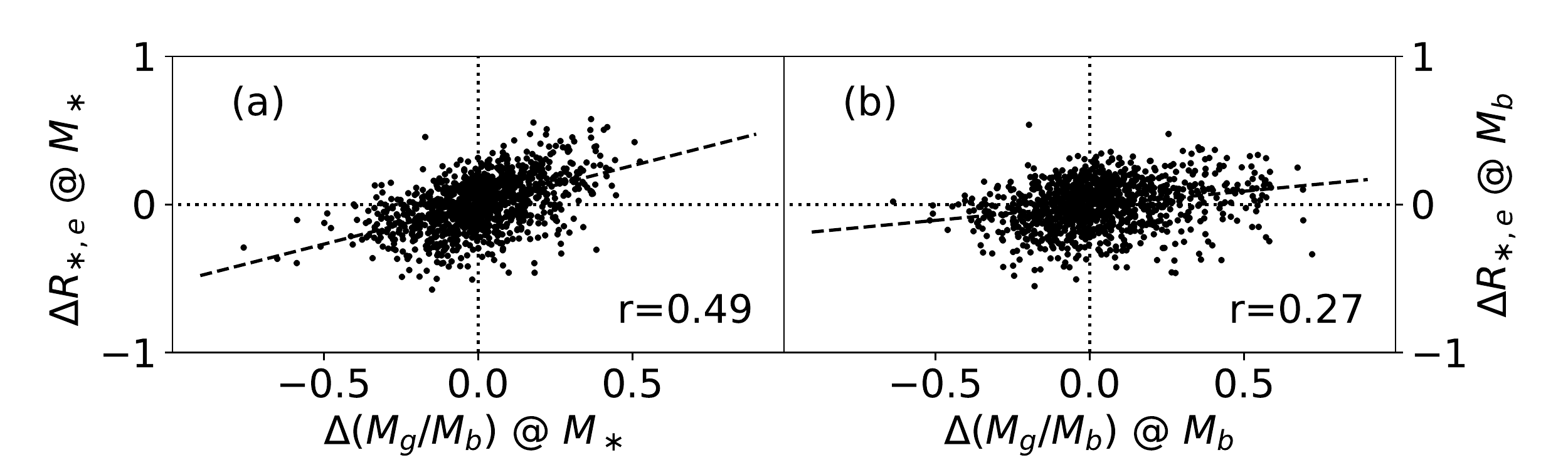}
	\caption{The correlation between the gas fraction and the effective radius of galaxies. At fixed stellar mass, gas-rich galaxies are on average larger, where at fixed baryonic mass, the correlation is weaker. Same as Fig.~\ref{fig:dfgdr}}
	\label{fig:dfgdre}
\end{figure*}

Fig.~\ref{fig:ReM}, \ref{fig:ReMfg}, and \ref{fig:dfgdre} are reproduction of Fig.~\ref{fig:RM}, \ref{fig:RMfg}, and \ref{fig:dfgdr}, using the effective radius $R_{\ast,e}$ as the definition of size instead of the scale length used in the main text. Overall, the conclusion remains the same if the effective radius is used.

The $R_{\ast,e}$-$M_b$ relation is a single power law with constant scatter at all masses and essentially no outlier. 
The best-fit $R_{\ast,e}$-$M_b$ relation is
\begin{equation}
\label{eq:mbre}
\log(R_{\ast,e}/kpc) = (0.342^{+0.013}_{-0.013}) \times [\log(M_b/M_\odot)-10] + (0.463^{+0.009}_{-0.009})
\end{equation}
The best-fit $R_{\ast,e}$--$M_\ast$ relation has a shallower slope of $0.279^{+0.011}_{-0.011}$ and an intercept of $0.517^{+0.009}_{-0.009}$.
With Equation~\ref{eq:mbre}, the size ratio between the stellar disc and the halo has a weak mass-dependence: $R_{\ast,e}/R_h \propto M_h^{0.16}$.

\section*{Acknowledgments}

I thank the anonymous referee for constructive and insightful comments. I also thank Dr. I-Ting Ho for his valuable opinion on the manuscript. 

\bibliographystyle{mnras}
\bibliography{MassSize}

\end{document}